\documentclass[apj]{emulateapj}  
 \usepackage{graphicx}
\usepackage{amsmath}
\usepackage{comment}
\usepackage{txfonts}
\usepackage{natbib}			

\renewcommand\email\texttt

\begin{document}
 
\slugcomment{\sc submitted to \it Astrophysical Journal}
\shorttitle{\sc A Bayesian Approach to locating the TRGB}
\shortauthors{Conn et al.}
 
\title{A Bayesian Approach to Locating the Red Giant Branch Tip Magnitude (Part I)}
 
\author{A.\ R. Conn\altaffilmark{1, 3}}

\author{G.\ F. Lewis\altaffilmark{2}}

\author{R.\ A. Ibata\altaffilmark{3}}

\author{Q.\ A. Parker\altaffilmark{1, 4}}

\author{D.\ B. Zucker\altaffilmark{1, 4}}

\author{A.\ W. McConnachie\altaffilmark{5}}

\author{N.\ F. Martin\altaffilmark{6}}

\author{M.\ J. Irwin\altaffilmark{7}}

\author{N. Tanvir\altaffilmark{8}}

\author{M.\ A. Fardal\altaffilmark{9}}

\author{A.\ M.\ N. Ferguson\altaffilmark{10}}
 
\altaffiltext{1}{Department of Physics \& Astronomy, Macquarie University, Sydney 2109, Australia }
 
\altaffiltext{2}{Institute of Astronomy, School of Physics, A29, University of Sydney, 
Sydney NSW 2006, Australia.} 
 
\altaffiltext{3}{Observatoire Astronomique, Universite« de Strasbourg, CNRS, 67000 Stras- 
bourg , France.}

\altaffiltext{4}{Australian Astronomical Observatory, PO Box 296, Epping, NSW 2121, Australia }

\altaffiltext{5}{NRC Herzberg Institute of Astrophysics, 5071 West Saanich Road, Victoria, British Columbia, Canada V9E 2E7}

\altaffiltext{6}{Max-Planck-Institut f{\"u}r Astronomie, K{\"o}nigstuhl 17, D-69117 Heidelberg, Germany}

\altaffiltext{7}{Institute of Astronomy, University of Cambridge, Madingley Road, Cambridge CB3 0HA, UK}

\altaffiltext{8}{Department of Physics and Astronomy, University of Leicester, Leicester LE1 7RH, UK}

\altaffiltext{9}{University of Massachusetts, Department of Astronomy, LGRT 619-E, 710 N. Pleasant Street, Amherst, Massachusetts 01003-9305, USA}

\altaffiltext{10}{Institute for Astronomy, University of Edinburgh, Royal Observatory, Blackford Hill, Edinburgh EH9 3HJ, UK}
 
\begin{abstract}
We present a new approach for identifying the Tip of the Red Giant Branch (TRGB) which, as we show, works robustly even on sparsely populated targets. Moreover, the approach is
highly adaptable to the available data for the stellar population under study, with prior information readily incorporable into the algorithm. The uncertainty in the derived distances is also made tangible and easily calculable from posterior probability distributions. We provide an outline of the development of the algorithm and present the results of tests designed to characterize its capabilities and limitations. We then apply the new algorithm to three M31 satellites: Andromeda I, Andromeda II and the fainter Andromeda XXIII, using data from the Pan-Andromeda Archaeological Survey (PAndAS), and derive their distances as $731^{(+ 5) + 18}_{(- 4) - 17}$ kpc, $634^{(+ 2) + 15}_{(- 2) - 14}$ kpc and $733^{(+ 13)+ 23}_{(- 11) - 22}$ kpc respectively, where the errors appearing in parentheses are the components intrinsic to the method, while the larger values give the errors after accounting for additional sources of error. These results agree well with the best distance determinations in the literature and provide the smallest uncertainties to date. This paper is an introduction to the workings and capabilities of our new approach in its basic form, while a follow-up paper shall make full use of the method's ability to incorporate priors and use the resulting algorithm to systematically obtain distances to all of M31's satellites identifiable in the PAndAS survey area.    
\end{abstract}
 
 
\section{Introduction}
 
The Tip of the Red Giant Branch is a very useful standard candle for gauging distances to extended, metal-poor structures. The tip corresponds to the very brightest members of the First Ascent Red Giant Branch (RGB), at which point stars are on the brink of fusing helium into carbon in their cores and hence contracting and dimming to become Horizontal Branch stars. The result is a truncation to the Red Giant Branch when the Color-Magnitude Diagram (CMD) for an old stellar population is generated, beyond which lie only the comparatively rare Asymptotic Giant Branch Stars and sources external to the system of interest. The (highly variable) contamination from such objects provides the principal obstacle to simply `reading off' the tip position from the RGB's luminosity function and the truncation of the AGB can even masquerade as the TRGB in certain instances. The I-band is the traditionally favored region of the spectrum for TRGB measurements, minimizing the interstellar reddening that plagues shorter wavelengths, while keeping dependance on metallicity lower than it would be at longer IR wavelengths. It should also be remembered that stars approaching the TRGB generally exhibit peak emission in this regime. \citet{Iben83} determined that low mass ($< 1.6 \, {\rm M}_{\odot}$ for Pop. I, $< 1 \, {\rm M}_{\odot}$ for Pop. II), metal-poor ([Fe/H] $<$ -0.7 dex) stars older than 2 Gyr produce a TRGB magnitude that varies by only 0.1 magnitudes. More recently, \citet{Bellazzini01} determined the tip magnitude to lie at an I-band magnitude of $M_{TRGB} = -4.04 \pm 0.12$. This low variation can be attributed to the fact that all such stars have a degenerate core at the onset of helium ignition and so their cores have similar properties regardless of the global properties of the stars. The result is a standard candle that is widely applicable to the old, metal-poor structures that occupy the halos of major galaxies. Distances derived from the TRGB, unlike those from the Cepheid variable or RR Lyrae star for example, can be determined from a single epoch of observation, making it very useful for wide-area survey data. Furthermore, \citet{Salaris97} confirmed agreement between Cepheid and RR Lyrae distances and TRGB distances to within $\sim$5 \%.   

Up until \citet{Lee93} published their edge-finding algorithm, the tip had always been found by eye, but clearly if the wide-reaching applications of the TRGB standard candle were to be realized, a more consistent, repeatable approach was in order. The aforementioned paper shows that, if a binned luminosity function (LF) for the desired field is convolved with a zero sum Sobel kernel [-2, 0, +2], a maximum is produced at the magnitude bin corresponding to the greatest discontinuity in star counts, which they attribute to the tip. Using this method, they were able to obtain accuracies of better than 0.2 magnitudes. \citet{Sakai96} set out to improve on this approach by replacing the binned LF and kernel with their smoothed equivalents. To do this, they equate each star with a Gaussian probability distribution whose FWHM is determined by the photometric error at the magnitude actually recorded for the star. Then rather than each star having to fall in a single bin, it contributes to all bins, but most strongly to the bin at the magnitude recorded for it and less so the further a bin is from that recorded magnitude, weighted by the photometric error. This is illustrated in equation \ref{e_Sakai}: 

\begin{equation}
       \Phi(m) = \sum^N_{i=1}\frac{1}{\sqrt{2\pi\sigma^2_i}}\exp{\left[-\frac{(m_i - m)^2}{2\sigma^2_i}\right]}
       \label{e_Sakai}
 \end{equation}
where $m$ is the magnitude of the bin in question and $m_i$ and $\sigma^2_i$ are the central magnitude and variance respectively of the gaussian probability distribution for the \emph{i}th star. This method halved the error associated with the non-smoothed version of the algorithm and an identical smoothing is hence justly incorporated into the model LF for our Bayesian approach. 

In a more recent variation on the Edge Detection methods, \citet{Madore09} once again applied a Sobel kernel, but fit to a luminosity function built from composite stellar magnitudes $T \equiv I - \beta [(V-I)_0 - 1.50]$ where $\beta$ is the slope of the TRGB as a function of color. This they argue results in a sharper output response from the filter, and allows all stars, regardless of color, to contribute equally to the derived tip position. \citet{Rizzi07} derived a value of $0.22 \pm 0.02$ for $\beta$ after a study of 5 nearby galaxies, and show that it is quite consistent from one galaxy to another.    

\citet{Mendez02} made a departure from the simple `edge-finding' algorithms above by adapting a maximum likelihood model fitting procedure into their technique. They point out that the luminosity function faint-ward of the tip is well modeled as a power law:

\begin{equation}
	L (m \ge m_{TRGB}) = 10^{a(m - m_{TRGB})} 
	\label{e_LF}
\end{equation}
where $m \geq m_{TRGB}$ and $a$ is fixed at 0.3.  They then ascribe the location of the tip to the magnitude at which this power law truncates - i.e. $m = m_{TRGB}$.  Bright-ward of the tip they assume a functional form:

\begin{equation}
	L (m < m_{TRGB}) = 10^{b(m - m_{TRGB})-c} 
\end{equation}
where $b$ is the slope of the power law bright-ward of the tip and $c$ is the magnitude of the step at the RGB tip.

Such a model, though simplistic, is robust against the strong Poisson noise that is inevitable in more sparsely populated LFs, making it a significant improvement over the previous, purely `edge-finding' methods.

\citet{Makarov06} follow in a similar vein, demonstrating the proven advantages of a Maximum Likelihood approach over simple Edge Detection techniques, despite a model dependance. Unlike \citet{Mendez02} however, they allow $a$ as a free parameter, arguing its notable variance from $0.3$ and importantly, they smooth their model LF using a photometric error function deduced from artificial star experiments. One shortcoming of both of these methods however, is that the most likely parameter values alone are obtained, without their respective distributions or representation of their dependance on the other parameters. Also, with regard to the background contamination, the RGB LF in fact sits on top of non-system stars in the field and so rather than model the background exclusively bright-ward of the TRGB, the truncated power law of Eq. \ref{e_LF} can be added onto some predefined function of the contamination. 

Arguably the most successful method developed so far has been that devised by \citet{McConn04}. It has been used to ascertain accurate distances to 17 members of the Local Group \citep{McConn05}. It combines aspects of both `edge-finding' and model fitting to zero in more accurately on the tip. They argue that as the precise shape of the luminosity function at the location of the tip is not known, a simple Sobel Kernel approach that assumes a sharp edge to the RGB, does not necessarily produce a maximum at the right location. They instead use a least-squares model fitting technique that fits to the LF in small windows searching for the portion best modeled by a simple slope function. This, they reason, marks the location of the steepest decline in star counts which is attributable to the tip location. This method is capable of finding the tip location accurate to better than 0.05 magnitudes, although is still susceptible to be thrown off by noise spikes in a poorly populated LF.

Despite the merits of previous methods such as these, none of them work particularly well when confronted with the high levels of Poisson noise that abound in the more poorly populated structures of galaxy halos. Furthermore, in such conditions as these where the offset between detected and true tip position will likely be at its greatest, it is of great use to have a full picture of likelihood space, as opposed to merely the determined, most probable value. This has lead us to develop a new, Bayesian approach to locating the TRGB, specifically, one that incorporates a Markov Chain Monte Carlo (MCMC) algorithm. As shall become apparent in the next section, such a method is very robust against noise spikes in the luminosity function and allows all prior knowledge about the system to be incorporated into the tip-finding process - something lacking in the previous approaches. Further to this, the MCMC provides for a remarkably simple, yet highly accurate error analysis. It also makes it possible to marginalize over parameters to provide posterior probability distributions (PPDs) of each parameter, or to obtain plots of the dependance of each parameter on every other. In \S \ref{s_MCMC_Method}, a detailed explanation of our approach and its limitations is given. \S \ref{ss_ANDI} introduces the method by applying the algorithm to one of M31's brightest dwarf spheroidals, Andromeda I. \S \ref{ss_errors} discusses the nature of systematic errors that apply to the method. \S \ref{ss_Initial_Tests} investigates the accuracy that the basic method (before addition of priors) is capable of given the number of stars populating the LF for the field and the strength of the non-RGB background while \S \ref{ss_Composite_LFs} deals with its performance when faced with a composite luminosity function. \S \ref{s_More_Distances} then applies our new approach to two additional M31 dwarf satellite galaxies and \S \ref{s_Conclusions} summarizes the advantages of the method and outlines the expected applicability of the method in the immediate future. 
 
\section{Method}
\label{s_MCMC_Method}

\subsection{The MCMC Method}
\label{ss_ANDI}
The Markov Chain Monte Carlo method is an iterative technique that, given some model and its associated parameters, rebuilds the model again and again with different values assigned to each parameter, in order that a model be found that is the best fit to the data at hand. It does this by comparing the likelihood of one model, built from newly proposed parameter values, being correct for the data, as opposed to the likelihood for the model built from the previously accepted set of model parameters. The MCMC then accepts or rejects the newly proposed parameter values weighted by the relative likelihoods of the current and proposed model parameter values. At every iteration of the MCMC, the currently accepted value of each parameter is stored so that the number of instances of each value occurring can be used to build a likelihood distribution histogram - which can be interpreted as a PPD - for each model parameter. Hence, the MCMC is a way of exploring the likelihood space of complicated models with many free parameters or possible priors imposed, where a pure maximum likelihood method would be quickly overwhelmed. With the PPD generated, the parameter values that produce the best fit model to the data can simply be read off from the peak of the PPD for each parameter. Similarly, the associated error can be ascertained from the specific shape of the distribution. A detailed description of the MCMC with worked examples can be found in \citet{Gregory05}.

\begin{figure}[htbp]
\begin{center}
\includegraphics[width = 0.35\textwidth,angle=-90]{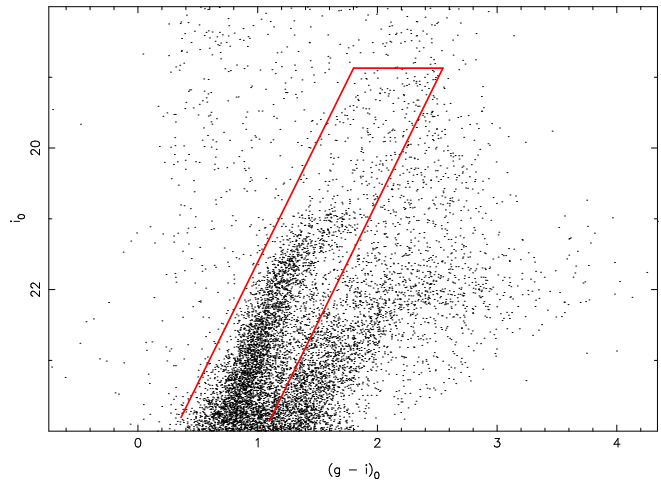}				  
\caption{Color-Magnitude Diagram for a circular field of radius $0.2^{\circ}$ centered on Andromeda I. Two red giant branches are clearly visible, that of Andromeda I (within the red rectangle color-cut) and that of the Giant Stellar Stream which lies behind Andromeda I in the same line of sight.}
\label{ANDI_CMD}
\end{center}

\end{figure}

To illustrate the precise workings of our MCMC tip-finding algorithm, its application to a well-populated dwarf  galaxy in the M31 halo is described. Andromeda I  was discovered by \citet{Bergh71} and at a projected distance of $\sim 45$ kpc from M31 \citep{Costa96}, it is one of its closest satellites. \citet{Costa96} ascribed to it an age of $\sim 10$ Gyr and a relatively low metallicity of $ \langle Fe/H \rangle = -1.45 \pm 0.2$ dex which is clearly exemplified in the color-magnitude diagram for Andromeda I presented in Fig \ref{ANDI_CMD}. Here the RGB of Andromeda I lies well to the blue side of that of the Giant Stellar Stream which lies behind Andromeda I but in the same field of view. \citet{Mould90} provide the first TRGB based distance measurement to Andromeda I, which they deduce as $790 \pm 60$ kpc, based solely on a visual study of the red giant branch. \citet{McConn04} improve on this significantly, producing a distance determination of $735 \pm 23$ kpc, based on a tip magnitude of $20.40^{+0.03}_{-0.02}$ in I band. 

\begin{figure}[htbp]
\begin{center}
\includegraphics[width = 0.55\textwidth,angle=-90]{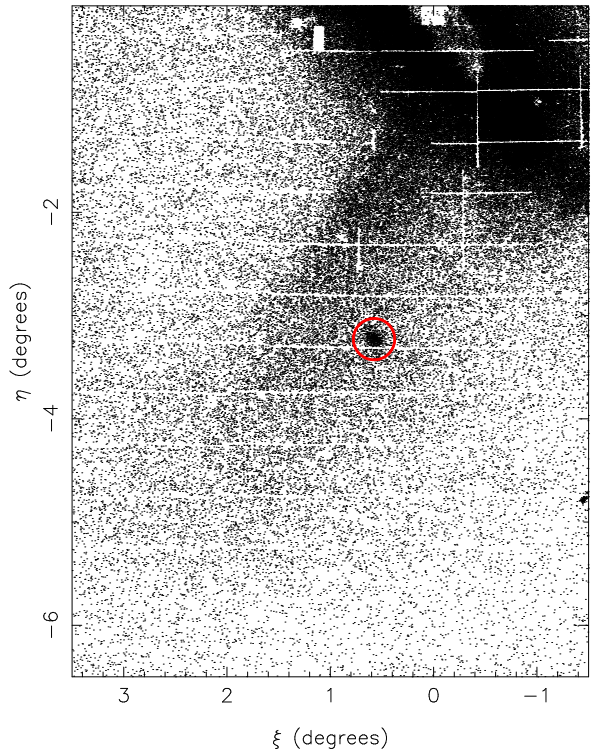}				  
\caption{The position of Andromeda I relative to the M31 disk. The saturated disk dominates the North West corner of the field whilst Andromeda I itself appears as an over-density within the Giant Stellar Stream (GSS). The GSS in actuality lies well behind Andromeda I, as is evidenced by the CMD in Fig. \ref{ANDI_CMD}. A strict color-cut was imposed on the data to highlight the location of the satellite and the extent of the stream with greatest contrast.}
\label{GSSandANDI}
\end{center}

\end{figure}

Andromeda I's position with respect to M31 and the Giant Stellar Stream is presented in Fig. \ref{GSSandANDI}, where the red circle indicates the precise field area fed to our MCMC algorithm. An Object-to-Background Ratio $(OBR)$ of $11.0$ was recorded for this field with the color-cut applied, based on comparisons of the signal field stellar density with that of an appropriate background field. The data presented in this figure, as with all other data discussed in this paper, was obtained as part of the Pan-Andromeda Archaeological Survey \citep{McConn09}, undertaken by the 3.6 m Canada-France-Hawaii Telescope (CFHT) on Mauna Kea equipped with the MegaCam imager. CFHT utilizes its own unique photometric band passes $i$ and $g$ based on the AB system. We work directly with the extinction-corrected CFHT $i$ and $g$ magnitudes and it is these that appear in all relevant subsequent figures. The extinction-correction data applied to each star has been interpolated using the data from \citet{Schlegel98}. 

At the heart of our tip-finding algorithm is the model luminosity function that the MCMC builds from the newly chosen parameters at every iteration. The luminosity function is a continuous function which we subsequently convolve with a Gaussian kernel to account for the photometric error at each magnitude. This is achieved by discretising both functions on a scale of 0.01 magnitudes.  Like \citet{Mendez02}, we assume the LF faint-ward of the tip to follow a simple power law, of the form given in Eq. \ref{e_LF}; however, we set $a$ as a free parameter. The bin height at each magnitude is then calculated by integrating along this function setting the bin edges as the limits of integration. The value for the bin which is set to contain the RGB tip for the current iteration is calculated by integrating along the function from the precise tip location to the faint edge of the bin. All other bins are then set at 0. A bin width of 0.01 magnitudes for our model was found to provide a good balance between magnitude resolution, which is limited by the photometric error in the MegaCam data ($\sim 0.01$ mag at m = 20.5), and the computational cost for a higher number of bins. We stress here however, that each star's likelihood is calculated from the model independently, so that the actual data LF is `fed' to the MCMC in an un-binned state. A faint edge to the model LF was imposed at $m = 23.5$ to remove any significant effects from data incompletion and increasing photometric error.

Further to this, we add a background function to this truncated power law. While the scaling of the background strength relative to the RGB signal strength could be set as another free parameter, and indeed was initially, it makes better use of our prior information to instead determine the fraction of background stars or `background height' ($f$) manually. This is achieved simply by calculating the average density of stars in the background field $D_{BG}$ and in the `signal' field $D_{SIG}$ with $f$ then being the ratio of the two, i.e. $f = D_{BG} / D_{SIG}$. Note that this is not directly the inverse of the object-to-background star ratio, $OBR = (D_{SIG} - D_{BG})/D_{BG}$, as $f$ represents the percentage of \emph{all} stars lying inside the signal field that can be expected to be external to the object of interest. Hence when we normalize the area under the model LF so that it may be used by the MCMC as a probability distribution, the background component will have area $f$ while the RGB component will have area $1 - f$. Now, with $f$ known, what we then have is a simplified 2 parameter model, allowing for faster convergence of the MCMC algorithm. 
  
We have thus devised our model so that the MCMC is tasked with the problem of finding just two parameters, namely the slope of the RGB luminosity function ($a$) and of course the location of the RGB tip magnitude ($m_{TRGB}$). For simplicity in this first paper, we impose uniform priors on each of these parameters, where $19.5 \le m_{TRGB} \le 23.5$ and $0 \le a \le 2$. We also do not account for the color dependence of the tip magnitude which is only slight in I band (see \citet{Rizzi07}) and for the metal poor targets examined here, but these effects will be dealt with in future publications. While it is true that two parameters are tractable analytically, we apply the numerical MCMC in order to set the framework for computationally more challenging models with non-uniform priors that will become necessary for the more sparsely populated structures presented in future contributions. There are however, several more complexities to the model that are yet to be discussed. Firstly, the choice of background function is not arbitrary. It has been found that the best way to model the background is to fit it directly by taking the luminosity function of an appropriate `background' field. The best choice of background field is arguably one that is at similar galactic latitude to the structure of interest, as field contamination is often largely Galactic in origin, and hence closely dependent on angular distance from the Galactic plane. Furthermore, the field should be chosen so that the presence of any substructure is minimal, so as to prevent the signature of another halo object interfering with the LF for the structure of interest. 

\begin{figure}[htbp]
\begin{center}
\includegraphics[width = 0.43\textwidth,angle=-90]{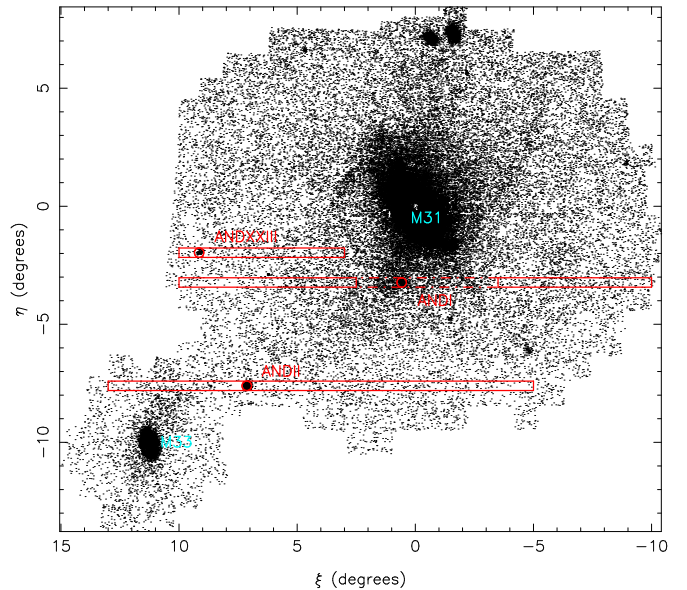}				  
\caption{A map of the entire PAndAS survey area, with color-cut chosen to favor the low metallicities exhibited by many of M31's satellite galaxies. The three dwarf spheroidal companions of M31 studied in this paper are labeled, along with the signal fields (small circles of radius $0.2^{\circ}$) and their respective background fields fed to our algorithm. Note that the background fields are chosen to be as narrow as possible in Galactic latitude while retaining as large an area as possible. In each case, the signal field areas are subtracted from their respective background fields to prevent contamination.}
\label{Sats_and_bckgrnds}
\end{center}

\end{figure}

In addition to these constraints, owing to the low stellar density of the uncontaminated halo, it is preferable that the background field be as large as possible to keep down the Poisson noise and hence it will of necessity be much larger than that of the field of interest. As a result, the main error in the background fit will arise from background mismatching and is not random. In addition, the large background field size may inevitably contain some substructure, requiring removal. This may be done by physically subtracting contaminated portions of the background area, but this is often unnecessary as the CMD color-cut imposed on the signal field must also be applied to the background field, usually ridding the sample of any substantial substructure that may be present. In the case of our Andromeda I background field however, we have removed a large $2.4^{\circ}$ portion crossing numerous streams (as shown in Fig. \ref{Sats_and_bckgrnds}) as these streams do trespass into the chosen Andromeda I color-cut. Nevertheless, this is just a precaution, because for well populated systems such as Andromeda I and Andromeda II, the algorithm is impervious to small discrepancies in the functional form of the background. 

\begin{figure}[htbp]
\begin{center} 
$ \begin{array}{c} 
\includegraphics[width = 0.35\textwidth,angle=-90]{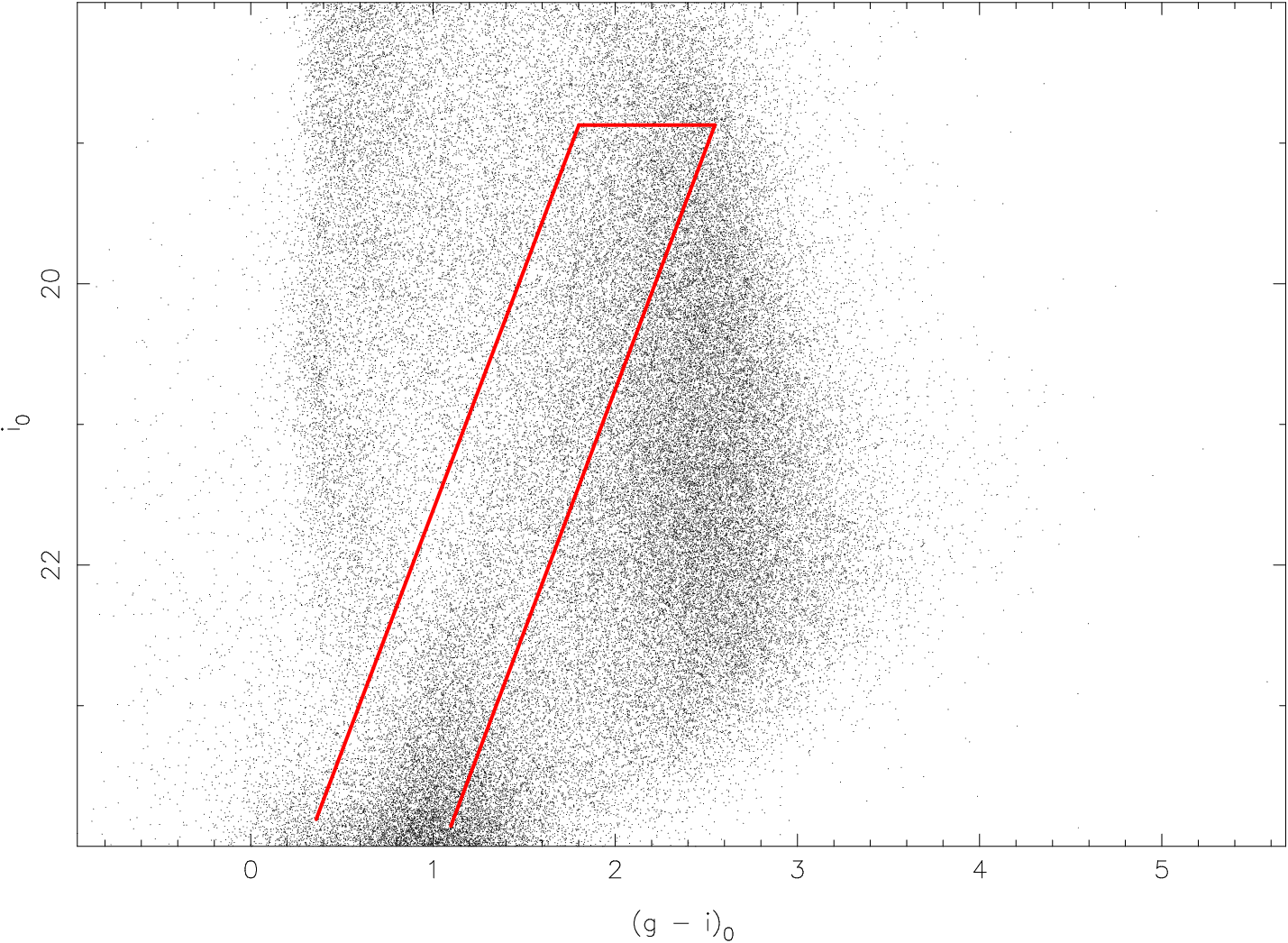}				
\\ \includegraphics[width = 0.35\textwidth,angle=-90]{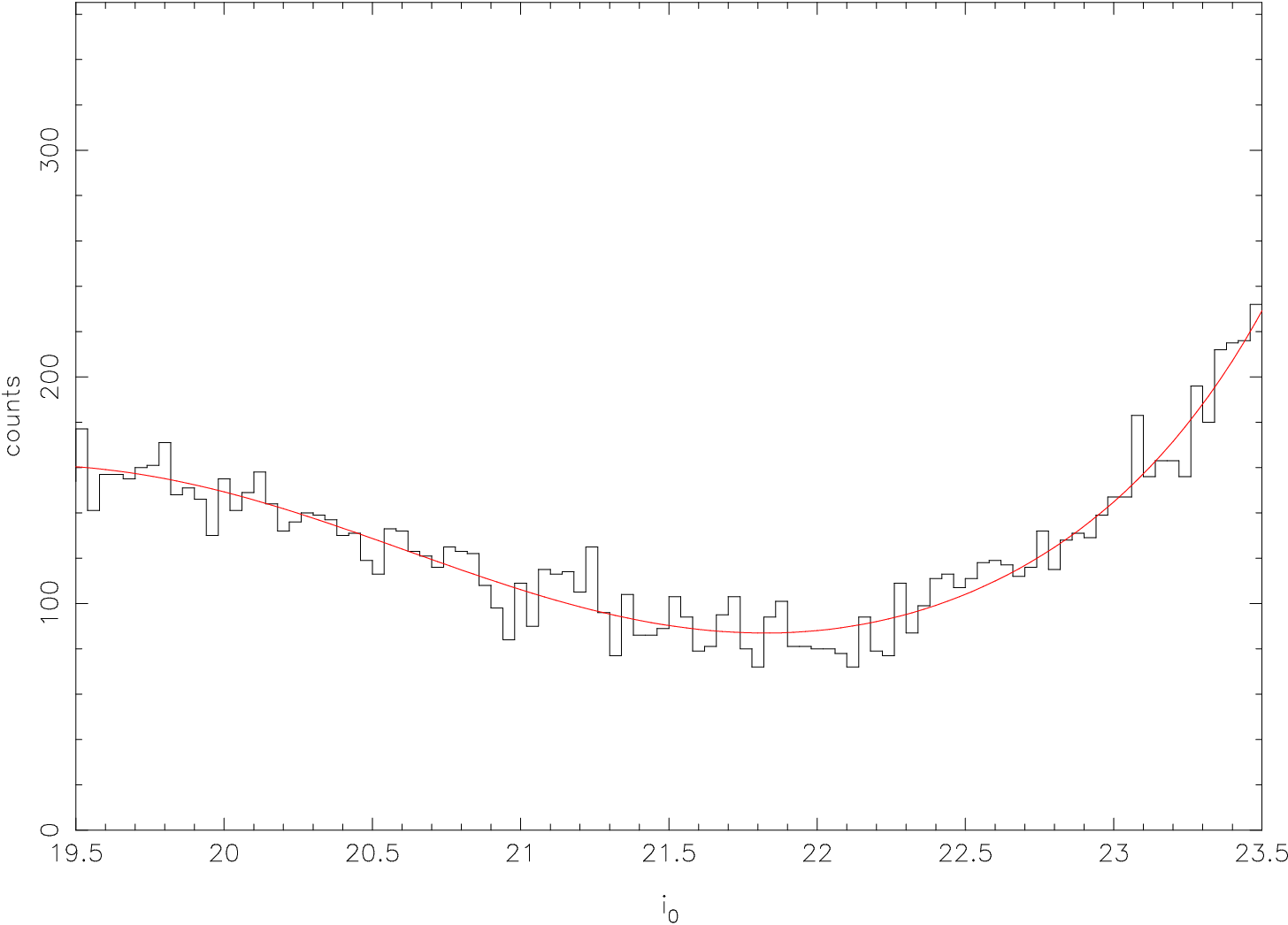} 
\end{array}$
\end{center}
\caption{Top) CMD for the Andromeda I background field (see Fig. \ref{Sats_and_bckgrnds}). The same color-cut is applied as in the CMD for the signal field (Fig. \ref{ANDI_CMD}). Bottom) The binned luminosity function for the background with the fitted polynomial superimposed. A polynomial of degree 7 was found adequate to represent the luminosity function.}
\label{GSS_Bckgrnd_LF}

\end{figure}

Once an appropriate background field has been selected, its LF can be fitted by a high-order polynomial. This polynomial then becomes the function added to our model and scaled by $f$ as described earlier. Our choice of background field for Andromeda I (along with Andromeda II and Andromeda XXIII) and the polynomial fit to its LF are presented in figures \ref{Sats_and_bckgrnds} and \ref{GSS_Bckgrnd_LF} respectively. 

The other major consideration that is yet to be addressed is the effect of photometric error on the luminosity function. This is dealt with by convolving the initial binned model with a normalized Gaussian whose width is adjusted as a function of magnitude in accordance with the error analysis conducted on the PAndAS data. This is equivalent to the method of \citet{Sakai96} described in Eq. \ref{e_Sakai}.  As described earlier, this procedure has the added advantage of making the model independent of binning. It is also important in this stage, as it is at every stage, that the model and all constituent parts are normalized so that the model can be used as a probability distribution.  

With these issues addressed, the MCMC algorithm can be set in motion. The $i$-band magnitudes and $(g-i)_0$ data for the desired field is read into data arrays, spurious sources are rejected and a color-cut imposed to remove as many non-members of the structure's RGB as possible. The same constraints are of course applied to the background field as well. The MCMC then applies pre-set starting values of $a$ and $m_{TRGB}$ and builds the corresponding model for the first iteration. Within this iteration, the MCMC proposes new values for each parameter, displaced by some random Gaussian deviate from the currently set values and re-constructs the appropriate model. The step size, or width of the Gaussian deviate is chosen so as to be large enough for the MCMC to explore the entire span of probability space, while small enough to provide a high resolution coverage of whatever features are present. The ratio of the likelihoods of the two models is then calculated (the Metropolis Ratio $r$) and a swap of accepted parameter values made if a new, uniform random deviate drawn from the interval [0,1], is less than or equal to $r$. The calculation of the Metropolis Ratio for our model is exemplified in equations \ref{e_MetroRat} and \ref{e_Likelihood}:

\begin{equation}
	r = \frac{{\cal L}_{proposed}}{{\cal L}_{current}}
	\label{e_MetroRat}
\end{equation}
with the value for each of the likelihoods ${\cal L}$ being calculated thus:

\begin{equation}
	{\cal L} = \prod_{n=1}^{ndata} M(m_{TRGB}, a, m_n)
	\label{e_Likelihood}
\end{equation}

with 

\begin{equation}
	 \begin{split}
	M(m_n \ge m_{TRGB}) &= RGB(m_n) + BG(m_n) \\
	M(m_n < m_{TRGB}) &= BG(m_n) \\
	where \; RGB(m_n) &= 10^{a(m_n - m_{TRGB})} \\
	and \; \int_{m = m_{TRGB}}^{m = 23.5} RGB \; dm &= 1 - f \\
	and \; \int_{m = 19.5}^{m = 23.5} BG \; dm &= f 
	 \end{split}
\end{equation}

where $m_{TRGB}$ and $a$ are the parameters currently chosen for the model by the MCMC, $ndata$ is the number of stars and $m_n$ is the $i$-band magnitude of the $n \,$th star. $BG$ represents the fitted background function (see Fig. \ref{GSS_Bckgrnd_LF}). The MCMC then stores the new choice for the current parameter values and cycles to the next iteration. In order to ascertain a reasonable number of iterations, the chains for each parameter were inspected to insure that they were well mixed, resulting in posterior distributions that appeared smooth (by eye).

\begin{figure}[htbp]
\begin{center}
\includegraphics[width = 0.35\textwidth,angle=-90]{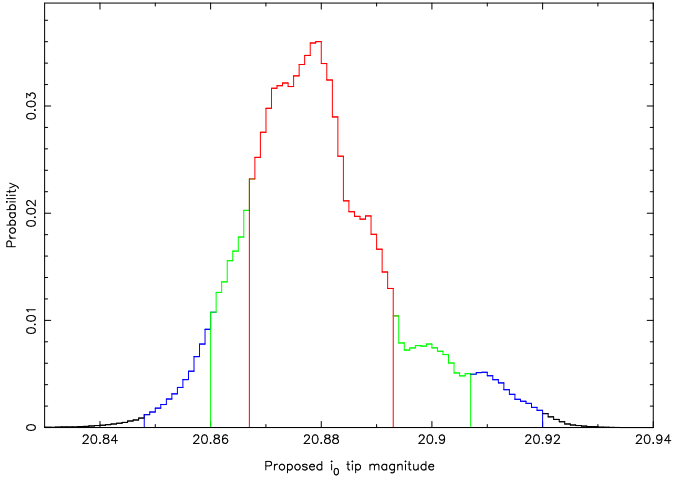}				  
\caption{The posterior probability distribution for 3 million iterations of the MCMC on the Andromeda I CMD color-cut presented in Fig \ref{ANDI_CMD}. The peak probability is located at $i_0$ = 20.88. The distribution is color coded, with red indicating tip magnitudes within 68.2 \% (Gaussian 1-sigma) on either side of the distribution mode, green those within 90 \% and blue those within 99 \%.}
\label{ANDI_PPD}
\end{center}

\end{figure}

\begin{figure}[htbp]
\begin{center}
\includegraphics[width = 0.35\textwidth,angle=-90]{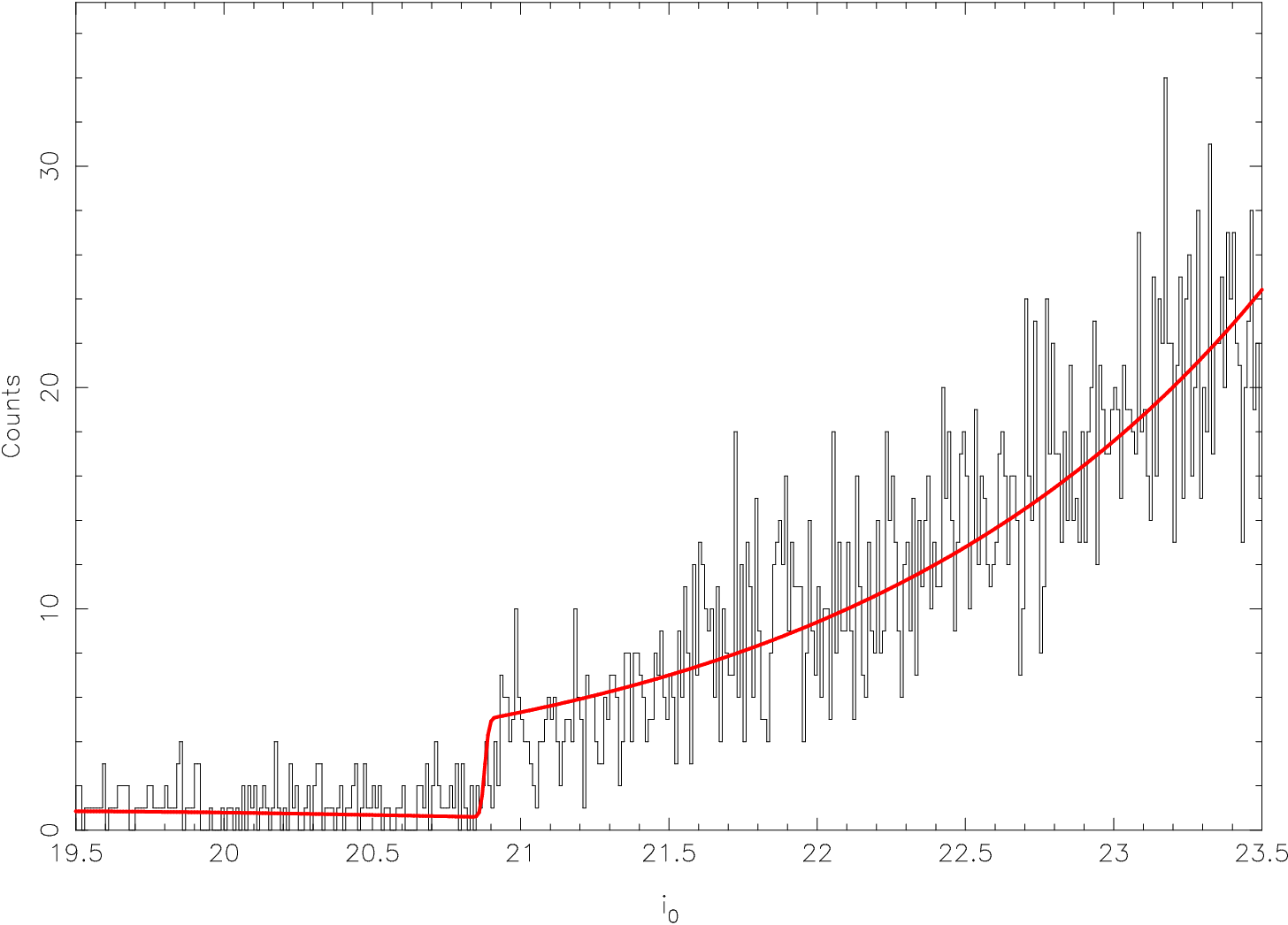}				  
\caption{The four magnitude segment of the Andromeda I luminosity function fitted by our MCMC algorithm. It is built from 3355 stars. The best fit model is overlaid in red. The bin width for the LF is 0.01 magnitudes.}
\label{ANDI_LF}
\end{center}

\end{figure}

\begin{figure}[htbp]
\begin{center}
\includegraphics[width = 0.35\textwidth,angle=-90]{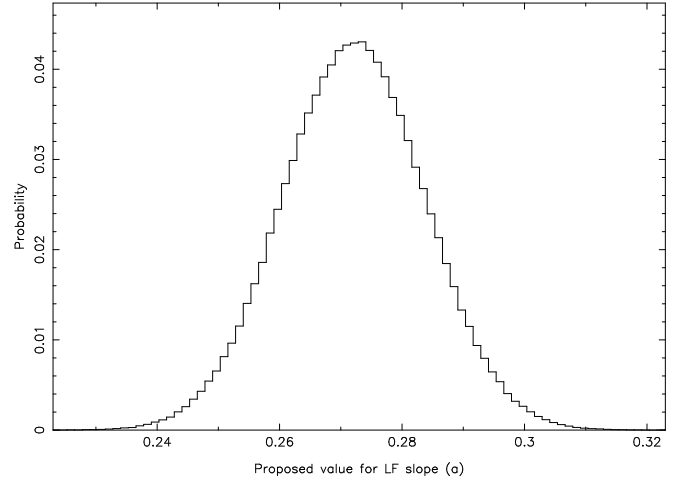}				  
\caption{The posterior probability distribution obtained for the slope $a$ of the Andromeda I luminosity function. The distribution is a clean Gaussian with the distribution mode at 0.273.}
\label{ANDI_a_PPD}
\end{center}

\end{figure}

\begin{figure}[htbp]
\begin{center}
\includegraphics[width = 0.35\textwidth,angle=-90]{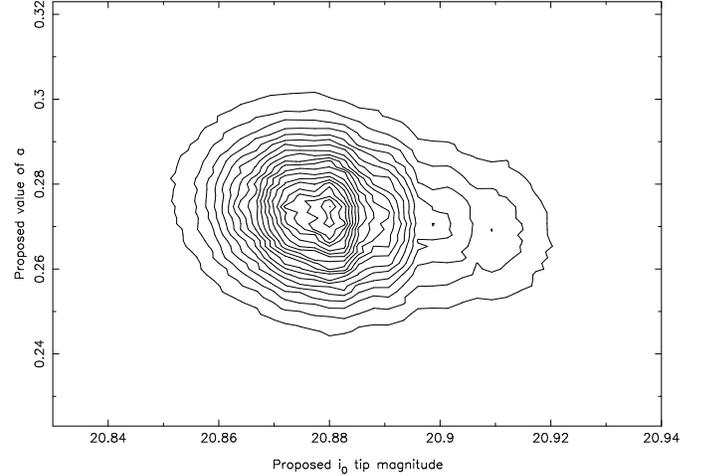}				  
\caption{A contour map of the distribution of the tip magnitude vs. the LF slope $a$. It is noteworthy that there is little correlation between the two parameters, with the peak of the distribution of $a$ more-or-less independent of tip magnitude. Regardless of any correlation, the respective PPDs of each parameter are the result of marginalizing over the other parameter, and thus take into account any covariance between parameters.}
\label{ANDI_a_vs_tip}
\end{center}

\end{figure}

When the MCMC has finished running, the posterior probability distribution (PPD) for each parameter is generated. By binning up the number of occurrences of each parameter value over the course of the MCMC's iterations, the probability of each value is directly determined and the most probable value can be adopted as the correct model value for the data. If one assumes a Gaussian probability distribution, then the 1-sigma errors associated with each parameter value can be obtained simply by finding the value range centred on the best-fit value that contains 68.2 \% of the data points. As our PPDs are not always Gaussian, our quoted 1-sigma errors in the tip magnitude represent more strictly a 68.2 \% credibility interval. We do not fit a Gaussian to our PPDs to obtain 1-sigma errors. Our 1-sigma errors in tip magnitude are obtained by finding the magnitude range spanning 68.2 \% of the PPD data points, on one side of the distribution mode and then the other. It must be stressed that these quoted errors are merely an indicator of the span of the parameter likelihood distribution and are no substitute for examining the PPDs themselves. Figures \ref{ANDI_PPD} and \ref{ANDI_LF} present the PPD for the RGB tip magnitude based on the Andromeda I CMD (Fig. \ref{ANDI_CMD}) and the best fit model to the LF for the field respectively. The PPD for the LF slope $a$ is presented in Fig. \ref{ANDI_a_PPD} and a contour map of the distribution of the tip magnitude vs. $a$ is presented in Fig. \ref{ANDI_a_vs_tip}.

Upon the completion of the algorithm, the RGB tip for Andromeda I was identified at $m = 20.879^{+ 0.014}_{- 0.012}$. This corresponds to an extinction-corrected distance of $731^{(+ 5) + 18}_{(- 4) - 17}$ kpc, where the final errors include contributions from the extinction and the uncertainty in the absolute magnitude of the TRGB (see \S \ref{ss_errors}). The $i$-band extinction in the direction of Andromeda I is taken as $A_{\lambda} = 0.105$ magnitudes \citep{Schlegel98}. The parameters $a$ and $f$ were derived as  $0.273 \pm 0.011$ and $0.083$ respectively. This distance measurement is in excellent agreement with the distance determined by \citet{McConn04}. It is noteworthy however that our method searches for the TRGB itself as distinct from the RGB star closest to the TRGB as sort out by the method of \citet{McConn04}, which would contribute to our slightly smaller distance measurement. A similar discrepancy arises in the case of Andromeda II (see \S \ref{s_More_Distances}).  

\subsection{A Note on Distance Errors}
\label{ss_errors}
Despite the small errors in the tip magnitude afforded by our approach, there are a number of factors that contribute to produce a somewhat larger error in the absolute distance. These arise due  to uncertainties both in the extinction corrections applied, and in the absolute magnitude of the TRGB in $i$-band.  Both of these contributions are assumed to be Gaussian, where the $1$-$\sigma$ error in the extinction correction, $ \Delta \{A_{\lambda}\}$, is taken as $10 \%$ of the correction applied, and the error in the absolute magnitude of the tip is expressed in equation \ref{e_dist_error} below:

 \begin{equation}
 \begin{split}
    \Delta \{M^{TRGB}_{i}\}& = \sqrt{\Delta^2\{m^{TRGB}_{i}\}_{\omega Cen} + \Delta^2\{A_{\lambda}\}_{\omega Cen} + \Delta^2\{m - M\}_{\omega Cen}}\\ 
    &= \sqrt{\{0.04\}^2 + \{0.03\}^2 + \{0.11\}^2}\\
    &= \pm 0.12
\end{split}
\label{e_dist_error}
\end{equation}

As we are working in the native CFHT $i$ and $g$ bands, we adopt this magnitude as $M^{TRGB}_{i} = -3.44 \pm 0.12$ where the conversion from $M^{TRGB}_{I}$ is based on the absolute magnitude for the TRGB identified for the SDSS i band \citep{Bellazzini08}. This is justified by the color equations applying to the new MegaCam i-band filter \citep{Gwyn10}. Noting that the largest contribution to this error is that from the distance modulus to $\omega \; Cen$, $(m - M)_{\omega Cen}$, derived from the eclipsing binary OGLEGC 17, we consider only the contributions from the extinction $\{A_{\lambda}\}_{\omega Cen}$, which is taken as $10\%$ of the \citet{Schlegel98} values, and the apparent tip magnitude determination $\{m^{TRGB}_{i}\}_{\omega Cen}$ and note that our derived distance modulus may be systematically displaced by up to 0.1 of a magnitude. This then gives us $M^{TRGB}_{i} = -3.44 \pm \sqrt{0.04^2 + 0.03^2} = -3.44 \pm 0.05$. Since our principal motive is to obtain relative distances between structures within the M31 halo rather than the absolute distances to the structures, this offset is not important. Furthermore, as measurements for the $\omega \; Cen$ distance modulus improve, our distances are instantly updatable by applying the necessary distance shift. 

\begin{figure}[htbp]
\begin{center}
\includegraphics[width = 0.35\textwidth,angle=-90]{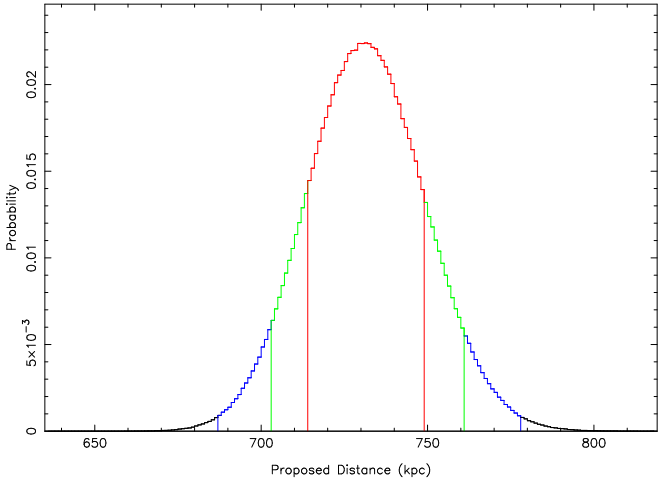}				  
\caption{A plot of the distribution of possible distances to Andromeda I obtained through the application of our method. Once again, the colors red, green and blue denote distances within $68.2 \%$, $90 \%$ and $99 \%$ credibility intervals respectively.}
\label{Dist_dist}
\end{center}

\end{figure}

While these external contributions to our distance uncertainties may be taken as Gaussian, the often non-Gaussian profile of our TRGB ($m_i^{TRGB}$) posterior distributions necessitate a more robust treatment then simply adding the separate error components in quadrature. Hence to obtain final distance uncertainties, we produce a Distance Distribution obtained by sampling combinations of $m_i^{TRGB}$, $A_{\lambda}$ and $M_i^{TRGB}$ from their respective likelihood distributions, thus giving us a true picture of the likelihood space for the distance. The result of this process for Andromeda I is illustrated in Fig. \ref{Dist_dist}.  From this distribution, we determine not only the quoted $1$-$\sigma$ errors, but also that Andromeda I lies at a distance between $703$ and $761$ kpc with $90 \%$ credibility and between $687$ and $778$ kpc with $99 \%$ credibility.   

\subsection{Initial Tests}
\label{ss_Initial_Tests}
In order to gain a better understanding of the capabilities of our method when faced with varying levels of  luminosity function quality, a series of tests were conducted on artificial `random realization' data, as well as on sub-samples of the Andromeda I field utilized above. There are two major factors that effect the quality of LF available to work with, namely the number of stars from which it is built, and the strength of the background component relative to the RGB component.  Hence to simulate the varying degrees of LF quality that are likely to be encountered in the M31 halo, artificial LFs were built for 99 combinations of background height vs. number of stars. Specifically, background heights of $f = 0.1, 0.2, ..., 0.9$ were tested against each of $ndata = 10, 20, 50, 100, 200, 500, 1000, 2000, 5000, 10000$ \& $20000$ stars populating the LF.

To achieve this, a model was built as discussed in \S \ref{ss_ANDI}, with a constant tip magnitude and RGB slope of $m_{TRGB} = 20.5$ and $ a = 0.3$ respectively and a background height $f$ set to one of the nine levels given above. The functional form of the background was kept as a horizontal line for the sake of the tests. A luminosity function was then built from the model, using one of the 11 possible values for the number of stars listed above. This was achieved by assigning to each of the $ndata$ stars a magnitude chosen at random, but weighted by the model LF probability distribution - a `random realization' of the model. The MCMC algorithm was then run on this artificial data set as described in the previous subsection with $m_{TRGB}$ and $a$ as free parameters to be recovered. The tests also assume the photometric errors of the PAndAS survey and further assume that incompleteness is not an issue in the magnitude range utilized. The error in the recovered tip position and the offset of this position from the known tip position in the artificial data ($I = 20.5$) were then recorded. The results are presented in Figures \ref{Sigma_Tests} \& \ref{Offset_Tests} below. Each pixel represents the average result of ten 200000 iteration MCMC runs for the given background height vs. number of stars combination. Note that the kpc distances given correlate to an object distance of $809$ kpc - i.e. $m^{TRGB}_I = 20.5$ - which is in keeping with distances to the central regions of the M31 halo. Furthermore, all stars of the random realization were generated within a 1 magnitude range centred on this tip value.

\begin{figure}[htbp]
\begin{center}
\includegraphics[width = 0.37\textwidth,angle=-90]{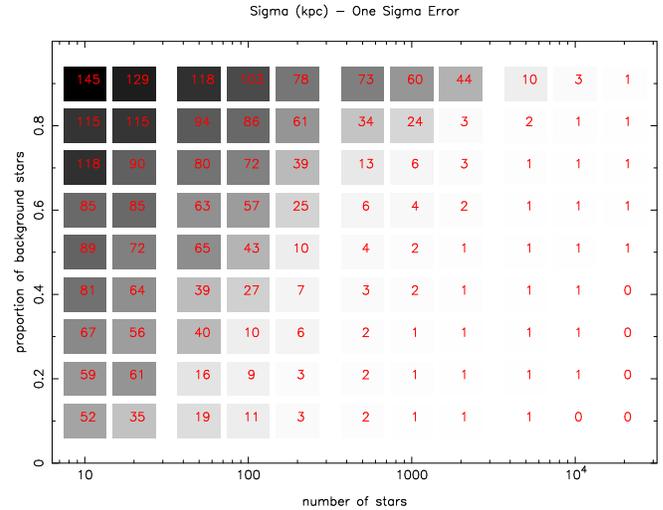}				  
\caption{A grey-scale map of the one-sigma error in tip magnitude obtained for different combinations of background height and number of sources. The actual value recorded for the error (in kpc) is overlaid on each pixel in red. For these tests, we approximate the one-sigma error as the half-width of the central 68.2 \% of the PPD span.}
\label{Sigma_Tests}
\end{center}

\end{figure}

\begin{figure}[htbp]
\begin{center}
\includegraphics[width = 0.37\textwidth,angle=-90]{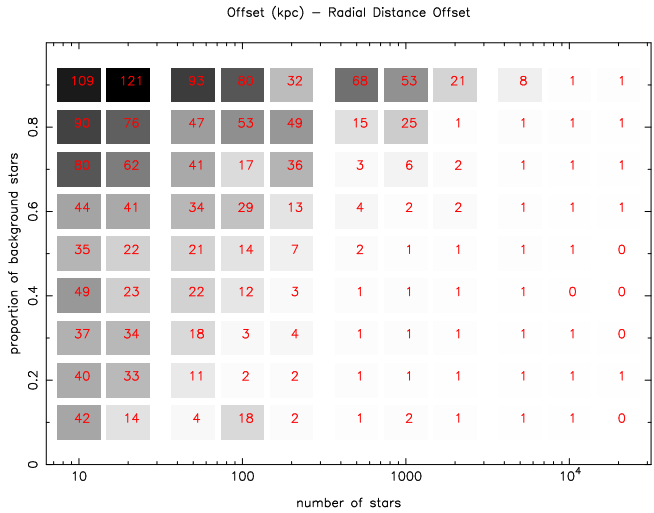}				  
\caption{A grey-scale map of the offset from the true tip magnitude obtained for different combinations of background height and number of sources. The actual (absolute) value recorded for the offset (in kpc) is overlaid on each pixel in red. These values convey the discrepancy between the true object distance and that recovered by the MCMC. It was necessary to remove the direction of the individual offsets before averaging as the values would otherwise largely cancel out. Examination of the individual offsets shows no significant bias toward either direction however.}
\label{Offset_Tests}
\end{center}

\end{figure}

Figures \ref{Sigma_Tests} \& \ref{Offset_Tests} are intended to serve as a reference for future use of the basic method, with regard to the number of stars required to obtain the distance to within the desired uncertainty for the available signal-to-noise ratio. The results follow the inevitable trend of greater performance when the background height is small and there are many stars populating the luminosity function. There are some minor deviations from this trend but these result from single outlying values whose effects would diminish if a higher number of samples were averaged. It is also noteworthy that the offsets recorded clearly correlate with the one-sigma errors and are consistently less than their associated errors.  

The results of these random realization tests are borne out by similar tests conducted on subsamples of the Andromeda I field. Random samples were drawn containing 335 (10\% of the total sample), 200, 100 and 50 stars. These correspond approximately to 10, 20, 50 and 100 stars in the 1 magnitude range centered on the tip. In no case was the derived tip location more than 80 kpc from that identified from the full sample, and the offset grew steadily less as the number of stars in the sample was increased. Furthermore, the offsets were almost always less than the 1-sigma errors. 

\subsection{Algorithm Behavior for Composite Luminosity Functions}
\label{ss_Composite_LFs}
When a field is fed to any RGB tip finding algorithm, it must be remembered that that field is in fact three dimensions of space projected onto two, and therefore it is possible that two structures at very different distances may be present within it. Such a scenario becomes especially likely when dealing with the busy hive of activity that the PAndAS Survey has come to reveal around M31. The result of such an alignment along the line of sight is a luminosity function built from two superimposed RGBs with two different - possibly widely separated - tip magnitudes. Hence it is important to understand how the TRGB algorithm applied to such a field will respond.

Unlike other algorithms that have been developed, our Bayesian approach provides us with a measure for the probability of the tip being at any given magnitude (the PPD). But this also leads to an important caveat - the selection criteria imposed on the data that is fed to the algorithm biases it strongly toward the structure whose distance we are trying to measure. Taking the Andromeda I measurement of \S \ref{ss_ANDI} for example, this satellite sits on top of the Giant Stellar Stream which contributes prominently to the field CMD, yet its contribution to the LF fed to the MCMC is almost eradicated by our choice of color-cut. Yet if this stringent color-cut is removed, the algorithm remains surprisingly insensitive to the GSS tip. This is because of another prior constraint we impose on the routine - the background height. With this fixed background imposed on our fitted model, the MCMC looks for the first consistent break of the data from the background - i.e. the tip of the Andromeda I RGB. It is therefore necessary to re-instate the background height as a free parameter of the MCMC to give it any chance of finding the tip of the Giant Stellar Stream's RGB. By this stage, enough of our prior constraints have been removed to give the method freedom to choose the best fit of the unrestricted model to the entire data set from the field. Nevertheless, the more (correct) prior information we can feed the algorithm, the better the result we can expect to receive.

Still, while the method has not been tailored towards composite luminosity functions, it is worth noting that it can be used successfully to identify more than one object in the line of sight - a useful ability when the two structures are poorly separated in color-magnitude space. The model used assumes only one RGB and thus one tip; to do otherwise would increase computation times. If two distinct structures are identified by this method and can not be separated using an appropriate color-cut or altered field boundaries, an appropriate double RGB model should be built to accurately locate the tip for each structure. But even with the basic single-RGB model (which will suffice for the vast majority of cases), at least the presence of a second structure is indicated.  If we take the example of Andromeda I again, the ideal way to obtain a distance measurement to the portion of the GSS that sits behind it would be to make a color-cut that favors it and removes Andromeda I, but we can force the algorithm to consider both structures to demonstrate the extreme case of what might be encountered in a general halo field. The result is two broad bumps in the PPD well separated in magnitude. The nature of the MCMC however is to converge straight onto the nearest major probability peak, seldom venturing far from that peak. This is remedied by the addition to the algorithm of Parallel Tempering.

\begin{figure}[htbp]
\begin{center}
\includegraphics[width = 0.35\textwidth,angle=-90]{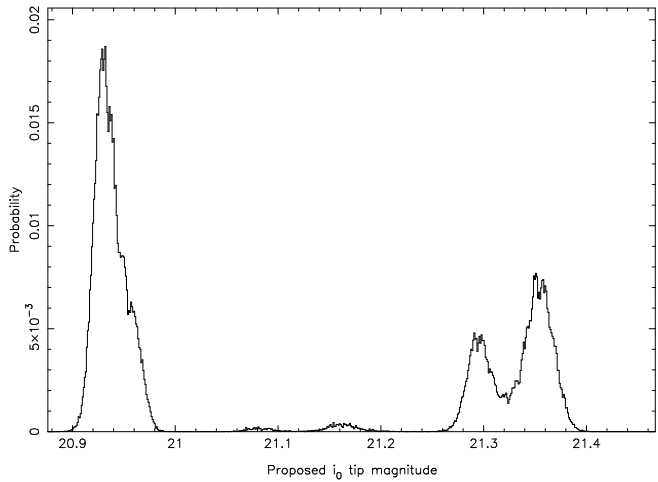}				  
\caption{The posterior probability distribution for the cold sampler chain of a 4 chain parallel tempering regime. The MCMC was run for 1.5 million iterations. The strong peak at $m = 20.93$ results from the tip of the Andromeda I RGB, but it has been shifted faint-ward by the presence of the Giant Stellar Stream, responsible for the peaks at $m = 21.29$ and $m = 21.35$. Without the addition of parallel tempering, the MCMC is liable to spend an inordinate amount of time stuck in the first major probability peak it encounters.}
\label{ANDI_PT_PPD}
\end{center}

\end{figure}

While an infinite number of iterations of the MCMC would accurately map probability space in its entirety, Parallel Tempering is a way of achieving this goal much more quickly. Parallel Tempering involves a simple modification to the MCMC algorithm, whereby multiple chains are run in parallel. One chain, the `cold sampler' runs exactly as before, but additional chains have their likelihoods weighted down producing a flatter PPD that is more readily traversed by the MCMC. The further the chain is from the cold sampler chain, the heavier the weight that is applied. Every so many iterations, a swap of parameters is proposed between two random, but adjacent chains so that even the `hottest' chains eventually affect the cold sampler chain and allow it to escape any local maximum it may be stuck in. The result is a cold sampler chain PPD that is more representative of the full extent of the luminosity function [see \citet{Gregory05} for a more detailed discussion]. The result of applying a 4 chain MCMC to the region of Andromeda I is summarized in the PPD of Fig. \ref{ANDI_PT_PPD}. 

While the Andromeda I TRGB is found much less accurately by this method as a result of the removal of our prior constraints for illustrative purposes, it is nevertheless clear that the addition of Parallel Tempering adds to our algorithm the facility to identify other structures in the field that may require separate analysis. Even given a properly constrained model and data set, the safeguard it provides against a poorly explored probability space arguably warrants its inclusion.

\section{Distances To Two More Satellites}
\label{s_More_Distances}
To further illustrate the capabilities of our basic method as outlined in \S \ref{s_MCMC_Method}, we have applied it to two more of M31's brighter satellites, whose distances have been determined in past measurements using a range of methods, including TRGB-finding algorithms. The additional satellites chosen for this study are the relatively luminous dwarf spheroidal Andromeda II and the somewhat fainter, newly discovered Andromeda XXIII dwarf. The location of both satellites within the M31 halo can be seen in Figure \ref{Sats_and_bckgrnds}. 
 
\subsection{Andromeda II} 
\begin{figure}[htbp]
\begin{center}
\includegraphics[width = 0.35\textwidth,angle=-90]{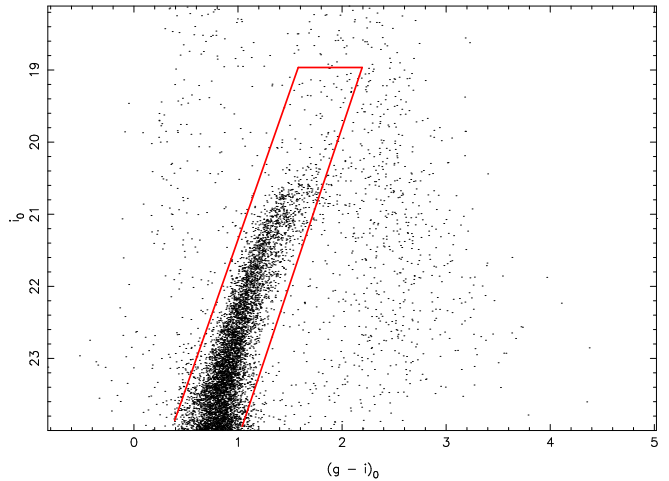}				  
\caption{Color-Magnitude Diagram for a circular field of radius $0.2^{\circ}$ centered on Andromeda II.  It is more densely populated than the Andromeda I CMD (Fig \ref{ANDI_CMD}) and is very well defined against the stellar background. The RGB tip is clearly visible at $i_0 \sim 20.6$.}
\label{ANDII_CMD}
\end{center}

\end{figure}

Andromeda II was discovered as a result of the same survey as Andromeda I using the 1.2 m Palomar Schmidt telescope \citep{Bergh71}. \citet{Costa00} deduce a similar age for Andromeda II as for Andromeda I but with a wider spread of metallicities centered on $ \langle Fe/H \rangle = -1.49 \pm 0.11$ dex. Our Andromeda II luminosity function was built from a circular field of radius $0.2^{\circ}$ centered on the dwarf spheroidal with an $OBR$ of $34.0$ recorded. This high OBR is not unexpected with Andromeda II arguably the best populated of M31's dwarf spheroidal satellites. The color magnitude diagram for this field is presented in Fig \ref{ANDII_CMD}.  

\begin{figure}[htbp]
\begin{center}
\includegraphics[width = 0.35\textwidth,angle=-90]{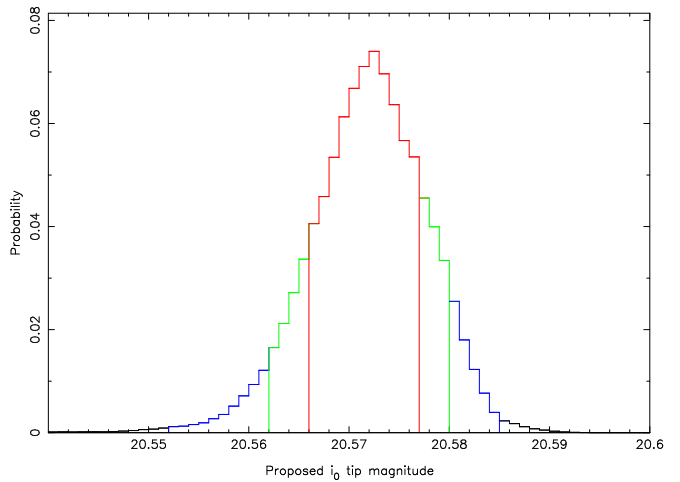}				  
\caption{The posterior probability distribution for 3 million iterations of the MCMC on a 4 magnitude interval (see Fig \ref{ANDII_LF}) of the Andromeda II CMD selection presented in Fig \ref{ANDII_CMD}. The peak probability of the distribution is well defined at $i_0 \approx 20.57$. The distribution is again color coded as in figure \ref{ANDI_PPD}, with red, green and blue corresponding to 68.2 \%, 90 \% and 99 \% credibility intervals respectively.}
\label{ANDII_PPD}
\end{center}

\end{figure}

\begin{figure}[htbp]
\begin{center}
\includegraphics[width = 0.35\textwidth,angle=-90]{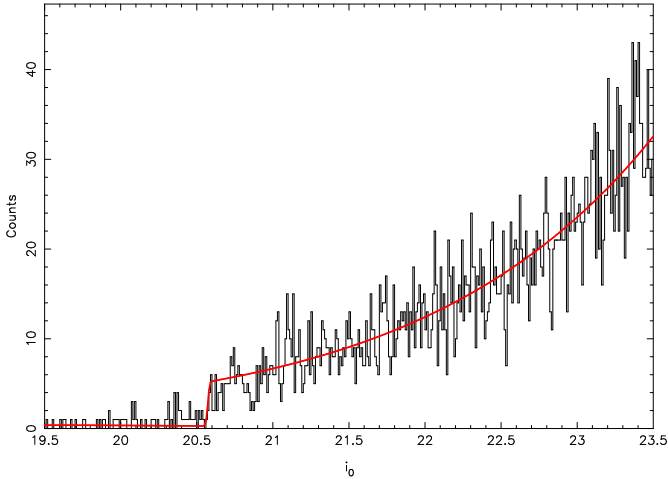}				  
\caption{The four magnitude segment of the Andromeda II luminosity function fitted by our MCMC algorithm. It is built from 4409 stars. The best fit model is overlaid in red. The bin width for the LF is again 0.01 magnitudes.}
\label{ANDII_LF}
\end{center}

\end{figure}

Application of our algorithm to Andromeda II yields a tip magnitude of $i_0 = 20.572^{+ 0.005}_{- 0.006}$ for the red giant branch which corresponds to an extinction-corrected distance to Andromeda II of $634^{(+ 2) + 15}_{(- 2) - 14}$ kpc, where the $i$-band extinction is taken as $A_{\lambda} = 0.121$ magnitudes \citep{Schlegel98}. This is in good agreement with \citet{McConn04}'s derived distance of $645 \pm 19$ kpc. Values for $a$ and $f$ were recovered as  $0.276 \pm 0.009$ and $0.028$ respectively. The $m_i^{TRGB}$ PPD and best fit model found by our method are illustrated in Figs. \ref{ANDII_PPD} and \ref{ANDII_LF} respectively. 

\subsection{Andromeda XXIII}

Despite its relative brightness among the other satellites of the M31 system, Andromeda XXIII was only discovered with the undertaking of the outer portion of the PAndAS survey in 2009/ 2010, being too faint at $M_V = -10.2 \pm 0.5$ to identify from the Sloan Digital Sky Survey \citep{Richardson11}. The said paper presents its vital statistics along with those for the other newly discovered satellites Andromeda XXIV - XXVII.  It is a dwarf spheroidal galaxy and has the lowest recorded metallicity of the satellites we present with $ \langle Fe/H \rangle = -1.8 \pm 0.2$. Making use of the deeper coverage of PAndAS in $g$-band, \citet{Richardson11} obtain a distance measurement of $767 \pm 44 $ kpc from the horizontal branch of the CMD.

\begin{figure}[htbp]
\begin{center}
\includegraphics[width = 0.35\textwidth,angle=-90]{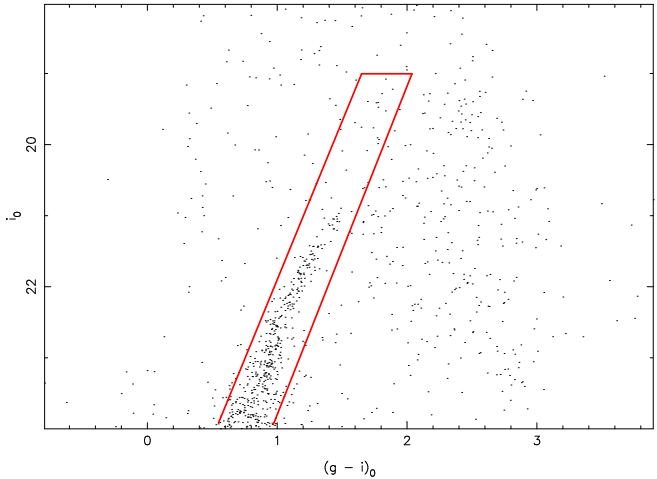}				  
\caption{Color-Magnitude Diagram for a circular field of radius $0.1^{\circ}$ centered on Andromeda XXIII.  It is much more sparsely populated than those of Andromeda I and Andromeda II. The RGB tip appears to lie just bright-ward of $i_0 = 21$.}
\label{ANDXXIII_CMD}
\end{center}

\end{figure}

\begin{figure}[htbp]
\begin{center}
\includegraphics[width = 0.35\textwidth,angle=-90]{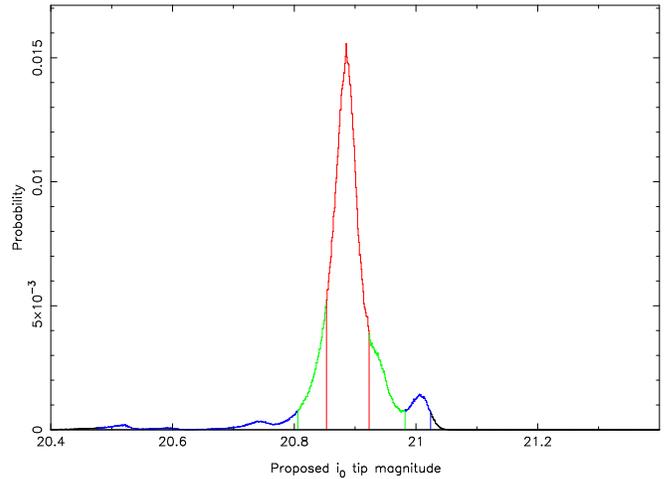}				   		
\caption{The posterior probability distribution for 3 million iterations of the MCMC on a 4 magnitude interval (see Fig \ref{ANDXXIII_LF}) of the Andromeda XXIII CMD selection presented in Fig \ref{ANDXXIII_CMD}. There are several probability peaks in this instance but the preferred peak lies at 20.885. The distribution is again color coded as in figure \ref{ANDI_PPD}, with red, green and blue corresponding to 68.2 \%, 90 \% and 99 \% credibility intervals respectively.}
\label{ANDXXIII_PPD}
\end{center}

\end{figure}

 Andromeda XXIII is a more challenging target for our algorithm in its current form, with less than $\sim 50$ stars lying within the one magnitude range centred on the tip and an $OBR$ of $8.4$ for the field and color-cut employed. The color-magnitude diagram for this circular field of radius $0.1^{\circ}$ is presented in Fig \ref{ANDXXIII_CMD}. We find the RGB tip at an $i$-band magnitude of $20.885^{+ 0.038}_{- 0.032}$, which, given an $i$-band extinction of 0.112 magnitudes in the direction of Andromeda XXIII \citep{Schlegel98}, corresponds to a distance of $733^{(+ 13)+ 23}_{(- 11) - 22}$ kpc. We derive the values of $a$ and $f$ as $0.270 \pm 0.039$ and $0.105$ respectively. Curiously, the MCMC finds several peaks very close to the major peak in the posterior probability distribution (see Fig \ref{ANDXXIII_PPD}) but these are attributable to the lower star counts available in the luminosity function around the tip. This has the effect of creating large magnitude gaps between the stars that are just brightward of the tip so that each individual star can mimic the sudden increase in star counts associated with the beginning of the RGB. As a result, there is a range of likely locations for the tip, but neglecting any error external to the method, the PPD shows that the object cannot be more distant than 867.0 kpc nor closer than 605.3 kpc with 99 \% confidence. 


\begin{figure}[htbp]
\begin{center}
\includegraphics[width = 0.35\textwidth,angle=-90]{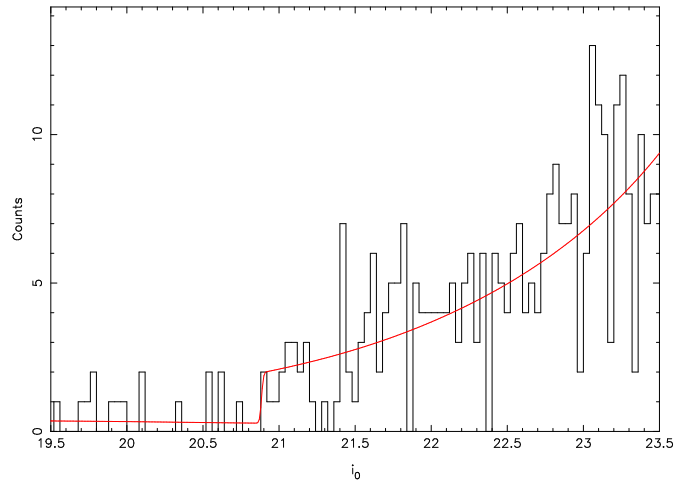}				  
\caption{The four magnitude segment of the Andromeda XXIII luminosity function fitted by our MCMC algorithm. It is built from 328 stars. The best fit model is overlaid in red. Whilst the model LF tested by the MCMC retained the resolution of 100 bins per magnitude described in \S \ref{ss_ANDI}, the data LF is re-produced here at the lower resolution of 0.04 magnitudes per bin to better reveal its structure to the eye.}
\label{ANDXXIII_LF}
\end{center}

\end{figure}

\section{Conclusions}
\label{s_Conclusions}

The versatility and robustness of our new method can be appreciated from \S \ref{s_MCMC_Method} and its high level of accuracy is evident from the measurement errors which are consistently smaller than those in the literature to date. In addition, it is our hope that with the correct priors imposed, this new approach carries with it the ability to gauge distances to even the most poorly populated substructures, bringing a whole new range of objects with in reach of the TRGB standard candle. In the case of the M31 halo alone, it will be possible to obtain distances to all of the new satellites discovered by the PAndAS survey - a feat previously impractical using the TRGB. Furthermore, PAndAS has revealed a complicated network of tidal streams that contain valuable information as to the distribution of dark matter within the M31 halo. With our new method, it will be possible to systematically obtain distances at multiple points along these streams, thus providing vital information for constraining their orbits. 

The great advantage of our new Bayesian method over a pure maximum likelihood method is the ease with which prior information may be built into the algorithm, making it more sensitive to the tip. Here in lies the great power of the Bayesian approach, whereby the addition of a few carefully chosen priors can reduce the measurement errors ten-fold. The result is an algorithm that is not only very accurate but highly adaptable and readily applicable to a wide range of structures within the distance (and metallicity) limitations of the TRGB standard candle. With instruments such as the 6.5 m Infra-red James Webb Space Telescope and the 42 m European Extremely Large Telescope expected to be operational within the decade, these distance limitations will soon be greatly reduced. This will bring an enormous volume of space within reach of the TRGB method, including the region of the Virgo Cluster. A tool with which it is possible to apply the TRGB standard candle to small, sparsely populated structures and small subsections of large structures alike is hence, needless to say, invaluable.

\begin{acknowledgments}
ACKNOWLEDGMENTS

 A. R. C. would like to thank Sydney University for allowing me the use of their computational and other resources. In addition, A. R. C. would like to thank fellow student Anjali Varghese, for her assistance with and practical insights with regard to the implementation of parallel tempering. A. R. C. would also like to thank Neil Conn for assistance in proof reading the document. 
G. F. L. thanks the Australian Research Council for support through his Future Fellowship (FT100100268) and Discovery Project (DP110100678). 
 
\end{acknowledgments}


\begin{thebibliography}{}

\bibitem[\protect\citeauthoryear{Bellazzini, Ferraro, 
\& Pancino}{2001}]{Bellazzini01} Bellazzini M., Ferraro F.~R., Pancino E., 2001, ApJ, 556, 635

\bibitem[\protect\citeauthoryear{Bellazzini
}{2008}]{Bellazzini08} Bellazzini M., 2008, Mem. S.A.It. Vol. 79, 440

\bibitem[\protect\citeauthoryear{Da Costa et 
al.}{1996}]{Costa96} Da Costa G.~S., Armandroff T.~E., Caldwell 
N., Seitzer P., 1996, AJ, 112, 2576

\bibitem[\protect\citeauthoryear{Da Costa et 
al.}{2000}]{Costa00} Da Costa G.~S., Armandroff T.~E., Caldwell 
N., Seitzer P., 2000, AJ, 119, 705 

\bibitem[\protect\citeauthoryear{Gregory 
}{2005}]{Gregory05} Gregory P.~C., 2005, Bayesian Logical Data Analysis for the Physical Sciences, Cambridge University Press, Ch. 12 

\bibitem[\protect\citeauthoryear{Gwyn
}{2010}]{Gwyn10} Gwyn, S., http://cadcwww.dao.nrc.ca/megapipe/docs/filters.html, 2010

\bibitem[\protect\citeauthoryear{Ibata et al.}{2007}]{Ibata07} 
Ibata R., Martin N.~F., Irwin M., Chapman S., Ferguson A.~M.~N., Lewis 
G.~F., McConnachie A.~W., 2007, ApJ, 671, 1591 

\bibitem[\protect\citeauthoryear{Iben 
\& Renzini}{1983}]{Iben83} Iben I., Jr., Renzini A., 1983, ARA\&A, 21, 271 

\bibitem[\protect\citeauthoryear{Lee, Freedman, 
\& Madore}{1993}]{Lee93} Lee M.~G., Freedman W.~L., Madore B.~F., 1993, ApJ, 417, 553 

\bibitem[\protect\citeauthoryear{Letarte et 
al.}{2009}]{Letarte09} Letarte B., et al., 2009, MNRAS, 400, 1472 

\bibitem[\protect\citeauthoryear{Madore, Mager, 
\& Freedman}{2009}]{Madore09} Madore B.~F., Mager V., Freedman W.~L., 2009, ApJ, 690, 389 

\bibitem[\protect\citeauthoryear{Makarov et 
al.}{2006}]{Makarov06} Makarov D., Makarova L., Rizzi L., Tully 
R.~B., Dolphin A.~E., Sakai S., Shaya E.~J., 2006, AJ, 132, 2729 

\bibitem[\protect\citeauthoryear{McConnachie et 
al.}{2003}]{McConn03} McConnachie A.~W., Irwin M.~J., Ibata 
R.~A., Ferguson A.~M.~N., Lewis G.~F., Tanvir N., 2003, MNRAS, 343, 1335 

\bibitem[\protect\citeauthoryear{McConnachie et 
al.}{2004}]{McConn04} McConnachie A.~W., Irwin M.~J., Ferguson 
A.~M.~N., Ibata R.~A., Lewis G.~F., Tanvir N., 2004, MNRAS, 350, 243 

\bibitem[\protect\citeauthoryear{McConnachie et 
al.}{2005}]{McConn05} McConnachie A.~W., Irwin M.~J., Ferguson 
A.~M.~N., Ibata R.~A., Lewis G.~F., Tanvir N., 2005, MNRAS, 356, 979 

\bibitem[\protect\citeauthoryear{McConnachie}{2009}]{McConn09} 
McConnachie A.~W., 2009, AAS, 41, 278 

\bibitem[\protect\citeauthoryear{M{\'e}ndez et 
al.}{2002}]{Mendez02} M{\'e}ndez B., Davis M., Moustakas J., 
Newman J., Madore B.~F., Freedman W.~L., 2002, AJ, 124, 213 

\bibitem[\protect\citeauthoryear{Mould 
\& Kristian}{1990}]{Mould90} Mould J., Kristian J., 1990, ApJ, 354, 438 

\bibitem[\protect\citeauthoryear{Richardson et 
al.}{2011}]{Richardson11} Richardson J.~C., et al., 2011, ApJ, 732, 
76

\bibitem[\protect\citeauthoryear{Rizzi et al.}{2007}]{Rizzi07} 
Rizzi L., Tully R.~B., Makarov D., Makarova L., Dolphin A.~E., Sakai S., 
Shaya E.~J., 2007, ApJ, 661, 815

\bibitem[\protect\citeauthoryear{Sakai, Madore, 
\& Freedman}{1996}]{Sakai96} Sakai S., Madore B.~F., Freedman W.~L., 1996, ApJ, 461, 713 

\bibitem[\protect\citeauthoryear{Salaris 
\& Cassisi}{1997}]{Salaris97} Salaris M., Cassisi S., 1997, MNRAS, 289, 406 

\bibitem[\protect\citeauthoryear{Schlegel, Finkbeiner, 
\& Davis}{1998}]{Schlegel98} Schlegel D.~J., Finkbeiner D.~P., Davis M., 1998, ApJ, 500, 525

\bibitem[\protect\citeauthoryear{van den Bergh}{1971}]{Bergh71} 
van den Bergh S., 1971, ApJ, 171, L31 

 
\end{thebibliography}
\end{document}